\documentclass[10pt]{article}
\input epsf

\begin{document}

\title{On a parity property in the thermal Sunyaev-Zel'dovich effect}
\small{\author{A. Sandoval-Villalbazo$^a$ and L.S. Garc\'{\i}a-Col\'{\i}n$^{b,\,c}$ \\
$^a$ Departamento de F\'{\i}sica y Matem\'{a}ticas, Universidad Iberoamericana \\
Lomas de Santa Fe 01210 M\'{e}xico D.F., M\'{e}xico \\
E-Mail: alfredo.sandoval@uia.mx \\
$^b$ Departamento de F\'{\i}sica, Universidad Aut\'{o}noma Metropolitana \\
M\'{e}xico D.F., 09340 M\'{e}xico \\
$^c$ El Colegio Nacional, Centro Hist\'{o}rico 06020 \\
M\'{e}xico D.F., M\'{e}xico \\
E-Mail: lgcs@xanum.uam.mx}} \maketitle

\bigskip
\begin{abstract}
{\small {The main issue in this paper is to discuss a parity
property that appears in the expressions for the distorted
spectrum of the thermal Sunyaev-Zel'dovich effect. When using the
convolution integrals method involving scattering laws we argue
that the distorted spectrum contains a new term, which is an odd
power of the frequency. Such a term, absent in the conventional
approaches, implies a crossover frequency which differs in value
form the ones reported in the literature by a significant, in
principle observable, amount. Also, such term casts doubt on the
demanding need of computing complicated relativistic calculations.
The relationship of our approach with the existing calculations is
discussed.}}
\end{abstract}

\section{Introduction}
The Sunyaev-Zel'dovich (SZ) effect is a signature left in the
cosmic microwave background (CMB) by collapsed structures
\cite{SZ1} \cite{SZ2}. The thermal Sunyaev-Zeldovich effect arises
from the frequency shift of CMB photons that are scattered by the
hot electrons contained in structures such as galaxy clusters. The
frequency dependence of this effect results, for a given line of
sight, in an intensity decrease in the Rayleigh-Jeans region of
the CMB spectrum and to an intensity increase at Wien's region.
The effect has been detected in observations of some rich, X-ray
luminous clusters. Besides this overall effect in the light
spectrum, of most interest is the value of the crossover
frequency, which is an observable quantity commonly used in the
determination of cosmological parameters \cite{Birk1}
\cite{Steen}. This observable is a strong ingredient in our
foregoing discussion.

The study of the non-relativistic and relativistic thermal effect
using analytical methods has led many authors to agree, one way or
another, upon the fact that the distorted spectra arising form
Compton scattering can be expressed as a power series containing
terms which are even powers of the frequency $\nu$, except for a
first  order in $\nu$ that is present due to the diffusion
approximation \cite{Sazonov} \cite{Stebbins} \cite{Itoh}
\cite{Challinor}. The parameter $z=\frac{k T_{e}}{m c^{2}}$ used
as a discriminant to indicate when the relativistic corrections
are important is usually kept up to second order effects. In this
paper, we challenge these results. We claim that if the correct
basic physics behind the inverse Compton scattering is used, a
term which is odd in $\nu$, in fact of order $z^{2 } \nu^{3}$ must
appear, leading to a couple of singular results. The first one is
that the crossover frequency, $\nu_{c}$, is subject to a $2.8$
percent correction with respect to the value predicted by
diffusion, in comparison to the $1.19$ percent correction obtained
with the relativistic equation. These values are taken for $k
T_{e}=5 KeV$. This could be presumably tested with precision
observations. The second implication is that, even when $k
T_{e}=15 KeV$, this odd power term, not only favors a value of
$\nu_{c}$ with an $8.4$ correction with respect to the $4.3$
percent predicted by relativistic corrections, but clearly
indicates that the curves for the distorted spectra are
practically indistinguishable from those obtained by numerical
methods, specially in Wien's region. Moreover, it is precisely in
this limit where claims have been made pointing out the opposite
\cite{Rephaeli}. In short, we sustain the old claim issued by
Sunyaev $25$ years ago \cite{Sun2}, stating that there is no need
to perform relativistic corrections to obtain the correct spectra
even at the upmost values for $k T_{e}=15 KeV$.

These arguments will be subsequently developed in sections two and
three of this paper.

\section{Gaussian scattering laws}

We begin this paper by recalling the approach followed previously
to discuss the SZ effect in terms of a scattering law \cite{nos1}
\cite{nos2}. This method is based upon the fact that the distorted
radiation spectrum which originates from inverse Compton
scattering between electrons and photons in a certain cluster is
given by:

\begin{equation}
I(\nu) = \int_{0}^{\infty} I_{o} (\bar{\nu}) G_{s} (\bar{\nu},\nu)
d \bar{\nu} \label{uno}
\end{equation}
Here, $I(\nu)$  is the scattered radiation off the plasma, $
I_{o}(\nu)=\frac{2 h \nu^{3}}{c^{2}}(exp(\frac{h \nu}{k
T_{R}})-1)^{-1}$ the undistorted spectrum, where $T_{R}$ is the
CMB temperature,  and $G_{s} (\bar{\nu},\nu)$ the scattering law.
For the thermal non-relativistic effect we have argued that the
form $ G_{s} (\bar{\nu},\nu)$ is given by \cite{nos1}

\begin{equation}
G_{s} (\bar{\nu},\nu) =(1-\tau) \delta (\bar{\nu}-\nu) +  \tau
G(\bar{\nu},\nu) \label{dos}
\end{equation}
where $\tau$ is the optical depth for a thin plasma, and
\begin{equation}
G(\bar{\nu},\nu)=\frac{1}{\sqrt{2 \pi }\sigma(\nu)}
e^{-(\frac{\bar{\nu}-(1-2z)\nu}{\sqrt{2} \sigma(\nu)})^{2}} .
\label{tres}
\end{equation}
Here, $\sigma^{2}(\nu)=2 \frac{k T_{e}} {m_{e} c^{2}} \nu^{2}=2 z
\nu^{2}$ is the square of the width of the spectral line at
frequency $\nu$, and $m_{e}$ and $T_{e}$ are the mass and
temperature of an electron and the electron gas, respectively.

The first term in Eq. (\ref{uno}) is self explanatory, it
represents the probability that a photon traverses the plasma
unscattered. The second term is a modified Gaussian probability
function describing the scattering of a photon with an incoming
frequency $\bar{\nu}$ by an electron whose average kinetic energy
is $\frac{1}{2} k T_{e}$. The reason of why the outgoing frequency
is shifted by a factor $2 z \nu $ arises precisely form the fact
that the average frequency of a photon scattered by electrons with
speeds $u=\beta c$ will exhibit a temperature blue shift given by
$2z \nu$ \cite{nos1} \cite{nos2}.

Inserting  Eqs. (\ref{dos}) and (\ref{tres}) into (\ref{uno}), and
upon integration  \cite{nos2}, one is lead to the result that

\begin{equation}
\frac{\Delta I(\nu)} {\tau} = -2z \nu \frac{\partial I_{o}}
{\partial \nu}+ z \nu^{2} \frac{\partial^{2} I_{o}} {\partial
\nu^{2}}+2 z^{2} \nu^{2} \frac{\partial^{2} I_{o}} {\partial
\nu^{2}}-2 z^{2} \nu^{3} \frac{\partial^{3} I_{o}} {\partial
\nu^{3}} + \frac{z^2 \nu^4} {2} \frac{\partial^{4} I_{o}}
{\partial \nu^{4}} \label{cuatro}
\end{equation}
to second order in $ z $ and first order in $\tau$. Eq.
(\ref{cuatro}) is a useful expression in order to discuss the SZ
thermal effect. Nevertheless, several characteristics of this
result must be underlined.  Here, of course, $\Delta
I(\nu)=I(\nu)-I_{o}(\nu)$.

Setting aside the fact that to first order in $z$ Eq.
(\ref{cuatro}) is identical to the diffusion approximation
described by the Kompaneets equation \cite{SZ1} \cite{SZ2}
\cite{seven}, a fact that has been analyzed in Ref. \cite{nos3},
we have other interesting features. The first one is the presence
of a third order derivative term in the right hand side of Eq.
(\ref{cuatro}) and the other one is that the last term leads to
the somewhat surprising fact that the Wien side of the distortion
curve is practically identical to its relativistic counterpart
\cite{nos2}. This is clearly exhibited in Fig.(1) for $k T_{e}=7.5
KeV $.
\begin{figure}
\epsfxsize=3.4in \epsfysize=2.6in
\epsffile{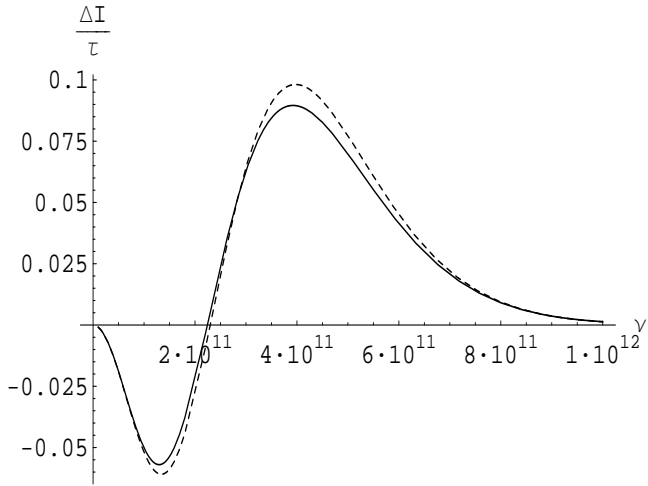}\vspace{0.5cm} \caption {\small{CMB
distortion for the case of Eq. (4) (dashed line)  compared with
the relativistic curve, Eq. (5) (solid line), here $k T_{e}=7.5
KeV$. Both curves are practically identical deep in the Wien
region. The frequency is given in Hz and $\frac{\Delta I}{\tau}$
is given in units of $\frac{2 (k T_{R})^{3}}{(h c)^{2}}$.} }
\end{figure}
\vspace{0.5cm}

A careful examination of the analytical expression for the
relativistic thermal SZ effect derived by Sazonov and Sunyaev
shows the absence of a third order derivative term. Indeed, Eq.
(12) of Ref. \cite{Sazonov} for a static cluster ($V=0$) is
identical to the expression:
\begin{equation}
\begin{array}{c}
\frac{\Delta I_{SS}}{\tau }=(-2\,z+\frac{17}{5}z^{2})\,\nu
\frac{\partial I_{o}}{
\partial \nu }+(\,z-\frac{17}{10}z^{2})\,\nu ^{2}\frac{\partial ^{2}I_{o}}{
\partial \nu ^{2}} \\
+\frac{7}{10}z^{2} \nu^{4} \frac{\partial ^{4}I_{o}}{\partial \nu
^{4}}
\end{array}
\label{cinco}
\end{equation}
a fact that can be easily proved by cumbersome but straightforward
algebra. Before attempting to analyze if a third order derivative
term should arise in the distortion curve, let us go back one step
and ask how is it possible to obtain an equation for $\frac{\Delta
I(\nu)} {\tau}$ without the third order derivative term. We don't
question the first derivative term, since it appears already in
the diffusion approximation.  The non-existence of the $\nu^{3}$
term can be achieved by writing the scattering law for $G_{s}$ as
follows:
\begin{equation}
G_{s}=(1-\tau) \delta (\bar{\nu}-(1-2z \tau) \nu) +  \tau
G'(\bar{\nu},\nu) \label{seis}
\end{equation}
where
\begin{equation}
G'(\bar{\nu},\nu)=\frac{1}{\sqrt{2 \pi } \sigma}
e^{-(\frac{\bar{\nu}-\nu} {\sqrt{2} \sigma})^2} \label{siete}
\end{equation}

Physically, this implies that $G'(\bar{\nu},\nu)$ is a purely
Gaussian function peaked at $\bar{\nu}=\nu$ \cite{Sun1} and that
the central limit theorem governs the scattering of photons by
electrons \cite{Ned}. Nevertheless, the equation would also imply
a blue shift in the frequency for the unscattered photons. In
fact, any even function $G'(\alpha)$ in the variable
$\alpha=\frac{\bar{\nu}-\nu} {\sqrt{2} \sigma}$ would suppress the
odd order higher derivative terms, as in Eq. (\ref{seis}).
Although we believe that this is physically incorrect,
substitution of Eqs. (\ref{seis}) and (\ref{siete}) into Eq.
(\ref{uno}) leads to an interesting result, namely,
\begin{equation}
\frac{\Delta I(\nu)_{a}}{\tau}=-2z \nu \frac{\partial
I_{o}}{\partial \nu}+ z \nu^{2} \frac{\partial^{2} I_{o}}{\partial
\nu^{2}}+\frac{1}{2} z^{2} \nu^{4}\frac{\partial^{4}
I_{o}}{\partial \nu^{4}} \label{ocho}
\end{equation}

In Fig. (2) we exhibit the effects of this result compared with
the Kompaneets approximation and with the full relativistic curve
of Sazonov and Sunyaev , Eq. (\ref{cinco}) \cite{Sazonov} for
$kT=10 KeV$. In Fig.(3) the curve arising form Eq. (\ref{cuatro})
is included. The reader may wonder why all this fuss about the
higher order in $z$ corrections but we believe that the underlying
physics is basic for  understanding the effect and, as we shall
see below, also to grasp the nature of the relativistic
corrections to $G_{s}$.
\begin{figure}
\epsfxsize=3.4in \epsfysize=2.6in
\epsffile{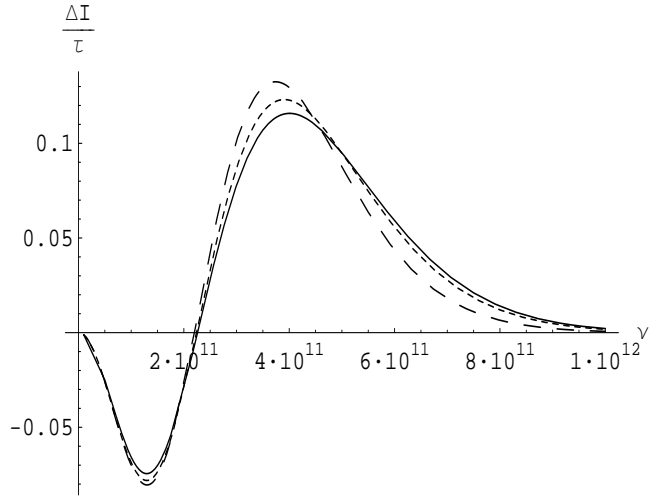}\vspace{0.5cm} \caption {\small{CMB
distortion for the case of Eq. (8) (short dashed line)  compared
with the Kompaneets approximation (long dashed line) and the
relativistic curve, Eq. (5) (solid line), $k T_{e}=10 KeV$. The
order $z^{2}$ curves are practically identical deep in  Wien's
region. The frequency is given in Hz and $\frac{\Delta I}{\tau}$
is given in units of $\frac{2 (k T_{R})^{3}}{(h c)^{2}}$.} }
\end{figure}
\vspace{0.5cm}

\begin{figure}
\epsfxsize=3.4in \epsfysize=2.6in
\epsffile{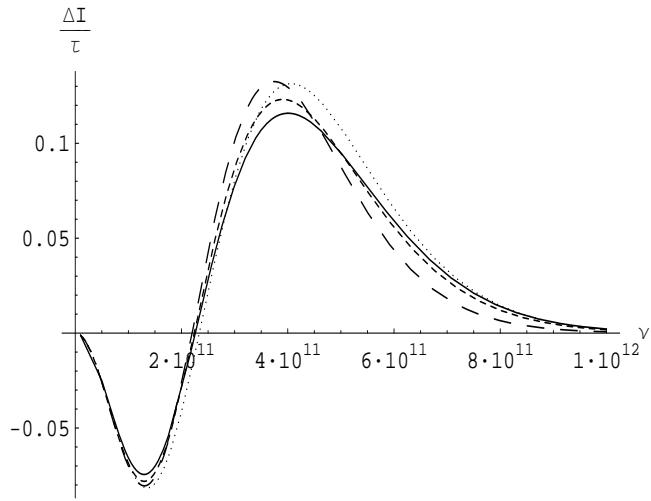}\vspace{0.5cm} \caption{The same as is Fig.
(2) now including a comparison with Eq. (4), (dotted line). The
third order derivative term induces an increase in the crossover
frequency and a \emph{better} agreement with the relativistic
curve at high frequencies.}
\end{figure}
\vspace{0.5cm}

Thus, we are brought back to the original question, namely, if
there is a third order derivative present in Eq. (\ref{cuatro}).
To obtain a guide for replying we went back to a detailed
examination of two papers, one by Stebbins written six years ago
\cite{Stebbins} and the publication of Sazonov and Sunyaev on
relativistic corrections  to both the thermal and kinematic SZ
effects \cite{Sazonov}. Although based on different approaches,
both led to results which turn out to be identical for the thermal
SZ effect. The same parity properties were found in related works
\cite{Stebbins} - \cite{Challinor}. Indeed, Eq.(10) of Ref.
\cite{Stebbins} is essentially an expression for the distorted
spectra written in terms of the occupation number
$n_{\gamma}=(e^{\frac{h \nu}{k T_{R}}}-1)^{-1}$ instead of the
intensity $I(\nu)$, noticing that
\begin{equation}\label{nueve}
\frac{\Delta n}{n_{o} \tau}=\frac{\Delta I}{I_{o} \tau}
\end{equation}
where $\tau$ is the optical depth, identifying $\epsilon=h \nu$
and assuming that $h \nu \ll m c^{2}$ in that equation. The
expansion of all terms of Eq.($10$) of Ref. \cite{Stebbins} is a
laborious but straightforward calculation that shows that the
third order derivative in $I_{o}$ disappears and that the
resulting expression is identical to Eq.(\ref{cinco}). The
surprising result  of the comparison between the non-relativistic
limit of Eq.(\ref{cinco}) and Eq. (\ref{cuatro}) is that the third
order derivative in $I_{o}$ ought to be absent. In our context,
this would imply that, for low temperature clusters $k T_{e}\simeq
5 KeV$, Eq. (\ref{ocho}) should be essentially correct, which we
object because it is physically questionable.

This leads us to a different test, namely the value of the
critical frequency $\nu_{c}$ defined by $\Delta I(\nu_{c})=0$. One
gets, by the Kompaneets equation that $\nu_{c}=220.027$ GHz, a
value which according to the relativistic correction increases,
for $k T_{e}=5 KeV$, only to $222.67$ GHz, a $1.19$ percent
correction. However, if the third order derivative is present in
Eq.(\ref{cuatro}), as we assert,   $\nu_{c}=226.39$ GHz, giving a
more significant correction of $2.81$ percent. The simple physics
behind Eq.(\ref{cuatro}) is quite attractive but merely
theoretical up to this point. Notice also that from the
observation of the results, it is really hard to see which is in
agreement with observations, all follow the same pattern,
specially in  Wien's region.

Finally, a comment on the last term of Eq.(\ref{cuatro}). Our
method, which is non-relativistic, yields a numerical factor of
$\frac{5}{10}$ instead of the $\frac{7}{10}$ factor obtained in
references \cite{Sazonov} and \cite{Stebbins} using explicit
relativistic corrections. The question is if this difference and
possible improvements can be obtained using the scattering law
approach in the relativistic case. We turn to this question in the
following section.
\bigskip

\section{Relativistic scattering law}
From the discussion of the previous section it is rather clear
that in order to introduce relativistic corrections into the
scattering Kernel $G_{s} (\bar{\nu},\nu)$ two facts must be
considered. First, the energy of an electron has to be written in
its relativistic form $E=m c^{2}(\gamma-1)$ and second, their
velocity distribution is no longer given by Maxwell's
distribution, but by its relativistic extension. Nevertheless,
since the relativistic parameter $z$ is small even at high
temperature clusters, $k T_{e}=15.1 KeV$, we may simply generalize
the scattering law, Eqs. (\ref{dos}) - (\ref{tres}), allowing for
$z^{2}$ corrections for the peak shift and $\sigma (\nu)$, namely
\begin{equation}
G_{r} (\bar{\nu},\nu) =(1-\tau) \delta (\bar{\nu}-\nu) +
\frac{1}{\sqrt{2 } N \sigma(\nu)} H(\frac{\bar{\nu}-(1-
\varepsilon)\nu}{\sqrt{2} \sigma(\nu)}) \label{diez}
\end{equation}
In this case, the evaluation of the expression $I(\nu) =
\int_{0}^{\infty} I_{o} (\nu) G_{r} (\bar{\nu},\nu) d \bar{\nu}$
leads to a generic Kompaneets type equation. This is achieved
performing the change of variable:
\begin{equation}
\alpha =\frac{\bar{\nu}-(1- \varepsilon)\nu}{\sqrt{2} \sigma(\nu)}
\label{once}
\end{equation}
with normalization factor
\begin{equation}
N =\int_{-\infty}^{\infty}H(\alpha) d \alpha \label{doce}
\end{equation}
and identifying
\begin{equation}
\Delta \nu(\alpha)=\sqrt{2} \sigma(\nu) \alpha - \epsilon \nu
\label{trece}
\end{equation}
The resulting distortion curve reads:
\begin{equation}
\begin{array}{c}
\frac{\Delta I_{r}}{\tau }=-\epsilon \nu \frac{\partial I_{o}}{
\partial \nu }+(\frac{\epsilon^{2}}{2})\,\nu ^{2}\frac{\partial ^{2}I_{o}}{
\partial \nu ^{2}} +(\frac{\sigma^{2}}{N})\frac{\partial ^{2}I_{o}}{
\partial \nu ^{2}} \int_{-\infty}^{\infty} \alpha^{2} H(\alpha) d \alpha\\
+(\frac{\sigma^{2} \epsilon \nu}{N})\frac{\partial ^{3}I_{o}}{
\partial \nu ^{3}} \int_{-\infty}^{\infty} \alpha^{2} H(\alpha) d \alpha+
(\frac{\sigma^{4}}{6 N}) \frac{\partial ^{4}I_{o}}{\partial \nu
^{4}} \int_{-\infty}^{\infty} \alpha^{4} H(\alpha) d \alpha
\end{array}
\label{catorce}
\end{equation}

Eq. (\ref{cuatro}) is a particular case of Eq. (\ref{catorce}) for
$\epsilon=-2z $, $\sigma^{2}(\nu)=2z \nu^{2}$ making
$H(\bar{\nu},\nu)=\frac{1}{\sqrt{2 \pi }\sigma(\nu)}
e^{-(\frac{\bar{\nu}-(1-2z)\nu}{\sqrt{2} \sigma(\nu)})^{2}}$. Of
course, the series can be trivially taken to arbitrary order in
$z^{n}$. It is interesting to notice that the coefficient of the
third order derivative term in Eq.\ref{catorce}) is independent of
relativistic effects up to second order in $z$ and would yield
$-2z \nu^{3}$ even taking into account relativistic corrections.
In Fig. (4) we exhibit the effects of Eq. (\ref{cuatro}) compared
with the Kompaneets approximation and with the full relativistic
curve of Sazonov and Sunyaev \cite{Sazonov} for $kT=15.1 KeV$. The
critical frequency shift correction will now increase from $4.347$
to $8.414$ percent, basically due to the presence of the third
derivative term.
\begin{figure}
\epsfxsize=3.4in \epsfysize=2.6in
\epsffile{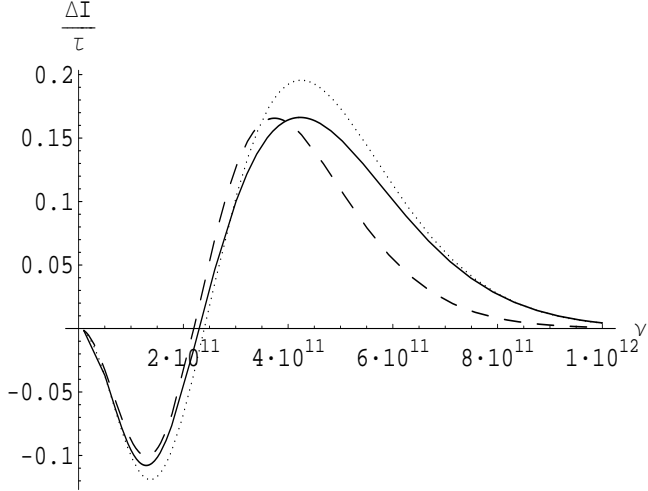}\vspace{0.5cm} \caption {\small{CMB
distortion for the case of Eq. (4) (solid line)  compared with Eq.
(5) (short dashed line) and with the Kompaneets approximation
(long dashed line) , $k T_{e}=15.1 KeV$. The crossover frequency
would have a large increase due to the presence of the third order
derivative term. The frequency is given in Hz and $\frac{\Delta
I}{\tau}$ is given in units of $\frac{2 (k T_{R})^{3}}{(h
c)^{2}}$.} }
\end{figure}
\vspace{0.5cm}

\section{Discussion}
The present paper not only shows that the $\nu^{3}$ term appearing
in Eq. (\ref{cuatro}) is essentially enough to account for the
relativistic features of the SZ thermal effect without entering
into long and sophisticated calculations, but provides a bridge
between the convolution integral approaches to the thermal SZ
effect and the expansions in terms of derivatives of the
undistorted number occupation . The calculation of convolution
integrals in the logarithmic space, which has been used by other
authors \cite{Birk1} \cite{Rephaeli}, prevents the establishment
of expressions such as Eq. (\ref{catorce})  that may relate the
approaches. On the other hand, the calculation of the convolution
integrals in the frequency space fills this gap. Similar
techniques applied to CMB distortions have been discussed in other
contexts \cite{Fixsen}.

To end this work we wish to underline the fact that the presence
of the third derivative term is directly directed to the frequency
shift included in the kernel of Eq. (\ref{tres}). If we take into
account that, on the average, a scattering produces a slight mean
change of photon energy \cite{Birk1} \cite{nos1}, then direct use
of the central limit theorem leads to Eq. (\ref{cuatro}).
\bigskip

This work has been supported by CONACyT (M\'{e}xico), project
$41081-F$.


\bigskip


\begin{thebibliography} {10}

\bibitem[1]{SZ1} Y.B. Zel'dovich and R.A. Sunyaev, Astrophys. Space Sci.
\textbf{4}, 301 (1969).

\bibitem[2]{SZ2} R.A. Sunyaev and Ya. B. Zel'dovich. Comm. Astrophys.
Space Phys. \textbf{4}, 173 (1972).

\bibitem[3]{Birk1}  M. Birkinshaw,  Phys.Rep. \textbf{310}
97 (1999) [astro-ph/9808050]

\bibitem[4]{Steen} A.D. Dolgov, S.H. Hansen, D.V. Semikoz and S. Pastor, Astrophys.
J. \textbf{ \ 554} 74 (2001) \ [astro-ph/0010412].

\bibitem[5]{Sazonov} S.Y. Sazonov and R.A. Sunyaev, Astrophys. J. \textbf{508},
1 (1998) [astro-ph/9804125].

\bibitem[6]{Stebbins} A. Stebbins (1997)  [astro-ph/9709065]

\bibitem[7]{Itoh}  N. Itoh, Y. Kohyama and S. Nozawa, Astrophys. J. \textbf{502},
7 (1998) [astro-ph/9712289].

\bibitem[8]{Challinor}  A. Challinor and A. Lasenby, Astrophys. J.
\textbf{499},1 (1998).  [astro-ph/9805329].


\bibitem[9]{Rephaeli} Y. Rephaeli, Astrophys. J. \textbf{445}, 33
(1995).

\bibitem[10]{Sun2} R.A. Sunyaev, Sov. Astrron. Lett. \textbf{6},
213 (1980).

\bibitem[10]{nos1} A. Sandoval-Villalbazo and L.S. Garc\'{\i}a-Col\'{\i}n;
J.Phys. A: Math and Gen. \textbf{36 }(2003) 4641-4650
[astro-ph/0208440].

\bibitem[11] {nos2} A. Sandoval-Villalbazo and L.S. Garc\'{\i}a-Col\'{\i}n;
Gen. Rel. and Grav. 2004 (in press) [astro-ph/0310465].

\bibitem[12]{seven} A.S. Kompaneets; JETP. \textbf{4}, 730 (1957).

\bibitem[13]{nos3} A. Sandoval-Villalbazo and L.S. Garc\'{\i}a-Col\'{\i}n (2003)
[astro-ph/0305144].


\bibitem[14]{Sun1} R.A. Sunyaev and Ya. B. Zel'dovich; Ann. Rev. Astrom.
Astrophys. \textbf{18}, 537 (1980)

\bibitem[15]{Ned} E.L. Wright, Astrophys. J. \textbf{232}, 348
(1979).

\bibitem[16] {Fixsen} R. Di Stefano, L.H. Ford, H. Yu and D.J. Fixsen
(2001) [astro-ph/0107001]



\end{thebibliography}
\end{document}